\documentclass[floatfix,aps,prl,showpacs,amsmath,nofootinbib,preprintnumbers,twocolumn]{revtex4}

\usepackage{graphicx}
\usepackage{bm}

\def\half{{1\over 2}}

\def\half{{1\over 2}}
\def\({\left(}
\def\){\right)}
\def\[{\left[}
\def\]{\right]}

\def\e{\begin{equation}}
\def\q{\end{equation}}
\def\m{\begin{eqnarray}}
\def\n{\end{eqnarray}}

\begin{document}

\title{Constraint on inflation model from BICEP2 and WMAP 9-year data}

\author{Cheng Cheng and Qing-Guo Huang}\email{huangqg@itp.ac.cn}
\affiliation{State Key Laboratory of Theoretical Physics, Institute of Theoretical Physics, Chinese Academy of Science, Beijing 100190, People's Republic of China}

\date{\today}

\begin{abstract}

Even though Planck data released in 2013 (P13) is not compatible with Background Imaging of Cosmic Extragalactic Polarization (B2) and some local cosmological observations, including Supernova Legacy Survey (SNLS) samples and $H_0$ prior from Hubble Space Telescope (HST) etc, Wilkinson Microwaves Anisotropy Probe 9-year data (W9) is consistent with all of them in the base six-parameter $\Lambda$CDM+tensor cosmology quite well. In this letter, we adopt the combinations of B2+W9 and B2+W9+SNLS+BAO+HST to constrain the cosmological parameters in the base six-parameter $\Lambda$CDM+tensor model with $n_t=-r/8$, where r and $n_t$ are the tensor-to-scalar ratio and the tilt of relic gravitational wave spectrum, and BAO denotes Baryon Acoustic Oscillation. We find that the Harrison-Zel'dovich (HZ) scale invariant scalar power spectrum is consistent with both data combinations, chaotic inflation is marginally disfavored by the data at around $2\sigma$ level, but the power-law inflation model and the inflation model with inverse power-law potential can fit the data nicely.

\end{abstract}


\maketitle


More than thirty years ago inflation \cite{Guth:1980zm,Linde:1981mu,Albrecht:1982wi} was proposed to solve the puzzles, such as the flatness problem, horizon problem, monopole problem and so on, in the hot big bang model. In fact, the spatial flatness can be taken as a prediction of inflation model which has been confirmed by Wilkinson Microwaves Anisotropy Probe 9-year data (W9) \cite{Hinshaw:2012aka} and Planck data released in 2013 (P13) \cite{Ade:2013zuv}. On the other hand, the quantum fluctuations generated during inflation \cite{Mukhanov:1981xt,Hawking:1982cz,Guth:1982ec,Starobinsky:1982ee,Bardeen:1983qw,Mukhanov:1985rz} can seed the anisotropies in the cosmic microwaves background (CMB) radiation and the formation of large-scale structure. Since the Hubble parameter during inflation is roughly a constant, the spectrum of scalar  perturbations is nearly scale-invariant. An adiabatic, Gaussian and nearly scale-independent scalar power spectrum has also been confirmed by W9 and P13. In addition, the quantization of the gravitational field during inflation produces a primordial background of stochastic gravitational waves \cite{Grishchuk:1974ny,Starobinsky:1979ty,Rubakov:1982,Barnaby:2012xt,Krauss:2013pha}. In the last decades, many group tried their best to hunt for the  signal of relic gravitational waves.

Recently discovery of relic gravitational waves was reported by Background Imaging of Cosmic Extragalactic Polarization (B2) \cite{Ade:2014xna}, and the tensor-to-scalar ratio is given by 
\m
r=0.20_{-0.05}^{+0.07}, 
\label{b2r}
\n
with $r=0$ disfavored at $7.0\sigma$. It is certainly a breakthrough of basic science in these years. Before B2 claimed its discovery, some hints of relic gravitational waves around $r\sim 0.2$ were illustrated in \cite{Zhao:2010ic,Zhao:2014rna} where only low-multipole CMB spectra are considered, and in \cite{Cheng:2013iya} from the combination of W9 \cite{Hinshaw:2012aka}, Atacama Cosmology Telescope (ACT) \cite{Sievers:2013ica}, South Pole Telescope (SPT) \cite{Story:2012wx}, Baryon Acoustic Oscillation (BAO) \cite{BAO} and $H_0$ prior from Hubble Space Telescope (HST) \cite{Riess:2011yx}.

Up to now, the $\Lambda$CDM model is widely accepted as a base model in which there are six parameters: baryon density today ($\Omega_b h^2$), cold dark matter density today ($\Omega_c h^2$), angular scale of the sound horizon at last-scattering ($\theta_{\rm MC}$), optical depth ($\tau$), scalar spectrum power-law index ($n_s$) and log power of the primordial curvature perturbations ($\ln (10^{10}A_s)$). Including the perturbations of primordial gravitational waves, the six-parameter $\Lambda$CDM model is extended to be $\Lambda$CDM+tensor model. However, combining with WMAP Polarization data \cite{Hinshaw:2012aka}, ACT \cite{Sievers:2013ica} and SPT \cite{Story:2012wx}, P13 \cite{Ade:2013zuv} imply a much smaller tensor-to-scalar ratio, compared to that from B2 in Eq.~(\ref{b2r}), 
\m
r<0.11 
\label{p13r}
\n
at $95\%$ C.L. in the base six-parameter $\Lambda$CDM+tensor cosmology. There is a moderately strong tension on $r$ between B2 and P13. Actually there are also several tensions between P13 and some local cosmological observations. For example, P13 prefers a larger matter density today compared to Supernova Legacy Survey (SNLS) samples \cite{Conley:2011ku}, and a smaller Hubble constant compared to the $H_0$ prior from HST \cite{Riess:2011yx}.

We noticed that W9 is consistent with almost all of other cosmological observations including B2, SNLS, BAO and $H_0$ prior from HST in the base six-parameter $\Lambda$CDM+tensor model. Therefore in this letter we constrain the cosmological parameters by respectively adopting the combinations of B2+W9 and B2+W9+SNLS+BAO+HST in the $\Lambda$CDM+r model where the tilt of relic gravitational waves spectrum is related to the tensor-to-scalar ratio by 
\m
n_t=-{r\over 8}
\n 
which is the consistency relation in the canonical single-field slow-roll inflation \cite{Liddle:1992wi}. For $r=0.2$, $n_t=-0.025$. A small value of $n_t$ is preferred by the data \cite{Cheng:2014bma,Cheng:2014ota}. Here the pivot scale is set as $k_p=0.004$ Mpc$^{-1}$.

\begin{figure}[hts]
\begin{center}
\includegraphics[width=3.4in]{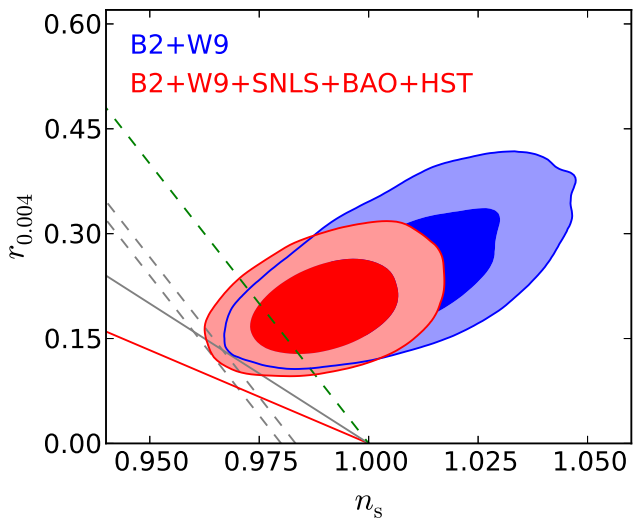}
\end{center}
\caption{The contour plot of $r$ and $n_s$ constrained by B2+W9 and B2+W9+SNLS+BAO+HST in the base $\Lambda$CDM+r model with $n_t=-r/8$. The blue and red contours correspond to B2+W9 and B2+W9+SNLS+BAO+HST respectively. The red solid line corresponds to inflation with $V(\phi)\sim \phi$. The gray solid line corresponds to $V(\phi)=\half m^2 \phi^2$, and the region between the two gray dashed lines corresponds to e-folding number within $N\in [50,\ 60]$. The green dashed line corresponds to the power-law inflation with potential $V(\phi)=V_0 \exp \(-\sqrt{2\over p}{\phi\over M_p}\)$.  
}
\label{fig:cssi}
\end{figure}

In the model of $\Lambda$CDM+r with $n_t=-r/8$, there are seven free running parameters, namely $\{\Omega_bh^2,\ \Omega_ch^2,\ \theta,\ \tau,\ n_s,\ A_s,\ r \}$. We run CosmoMC \cite{cosmomc} to fit all of these seven parameters by adopting the combinations of B2+W9 and B2+W9+SNLS+BAO+HST respectively. Our results are summarized in Table \ref{tab:cssi} and Fig.~\ref{fig:cssi}. 
\begin{table}[htbp]
\centering
\renewcommand{\arraystretch}{1.5}
\scriptsize 
{
 
\

\begin{tabular}{c|c|c}
\hline\hline
$\Lambda$CDM+r ($n_t=-r/8)$& \multicolumn{1}{|c|}{B2+W9} & \multicolumn{1}{c}{B2+W9+SNLS+BAO+HST} \\
\hline
parameters&$68\%$ limits & $68\%$ limits  \\
\hline
$\Omega_b h^2$ & $0.0236\pm 0.0006$ & $0.0231\pm 0.0004$\\
$\Omega_c h^2$ &  $0.1068_{-0.0045}^{+0.0046}$ & $0.1143\pm 0.0023$\\
100$\theta_{\rm MC}$ &  $1.0426\pm 0.0023$ & $1.0410\pm 0.0020$ \\
$\tau$ & $0.0932_{-0.0143}^{+0.0139}$ & $0.0868_{-0.0141}^{+0.0127}$ \\
$\ln(10^{10}A_s)$ & $3.070\pm 0.044$ & $3.122\pm 0.031$ \\
$n_s$ & $1.008_{-0.016}^{+0.015}$& $0.991_{-0.011}^{+0.010}$ \\
$r$& $0.25_{-0.08}^{+0.04}$&  $0.20_{-0.05}^{+0.04}$ \\
\hline
\end{tabular}
}
\caption{Constraints on the cosmological parameters from the combinations of B2+W9 and B2+W9+SNLS+BAO+HST in the $\Lambda$CDM+r model with $n_t=-r/8$ respectively.  }
\label{tab:cssi}
\end{table}
W9 is consistent with B2 quite well and we do not need to add any complicated physics, e.g. the running of spectral index. Even though P13 \cite{Ade:2013zuv} implies that the Harrison-Zel'dovich (HZ) scale invariant scalar power spectrum $(n_s=1)$ is disfavored at more than $5\sigma$, the HZ scalar power spectrum is consistent with the combinations of both B2+W9 and B2+W9+SNLS+BAO+HST nicely. In addition, since the contour plot of $r$ and $n_s$ in Fig.~\ref{fig:cssi} is above the red solid line corresponding to $V(\phi)\sim \phi$, it indicates that a convex potential of inflation field is preferred at more than $2\sigma$ level.

It is well-known that the chaotic inflation \cite{Linde:1983gd} proposed by A.~Linde can generate large amplitude of relic gravitational waves. The potential of inflaton field in the chaotic inflation is given by 
\m 
V(\phi)\sim \phi^n.
\n 
This model predicts 
\m 
r={4n\over N},
\n 
and 
\m
n_s=1-{n+2\over 2N},  
\n 
where $N$ is the number of e-folds before the end of inflation. Usually the CMB scales correspond to $N\simeq 50\sim 60$. For example, for $n=2$ and $N=50$, $r=0.16$ and $n_s=0.96$. In Fig.~\ref{fig:cssi}, the gray solid line corresponds to the prediction of chaotic inflation model with potential $\half m^2 \phi^2$ which is marginally disfavored at around $2\sigma$ level. 

In string theory, a general mechanism for chaotic inflation is proposed to be driven by monodromy-extended closed-string axion. See, for example, $n=2/3$ in \cite{Silverstein:2008sg} and $n=1$ in \cite{McAllister:2008hb}. For $n=2/3$ and $N=50$, $r=0.053$ and $n_s=0.973$; for $n=1$ and $N=50$, $r=0.08$ and $n_s=0.97$. Compared to the constraints on $r$ and $n_s$ in Fig.~\ref{fig:cssi}, the monodromy axion inflation models are disfavored at more than $2\sigma$ level. 

In fact, the the region between two gray dashed lines corresponds to the inflation with $V(\phi)\sim \phi^n$ where the e-folding number is $N\simeq 50\sim 60$. Because the $2\sigma$ region of $r-n_t$ in Fig.~\ref{fig:cssi} is almost above the region between two gray dashed lines, the chaotic inflation model is disfavored at around $2\sigma$ level.

Another well-known inflation model is the so-called power law inflation \cite{Lucchin:1984yf} which is govern by the potential  
\m 
V(\phi)=V_0 \exp \(-\sqrt{2\over p}{\phi\over M_p}\). 
\n
The spectral index and the tensor-to-scalar ratio in power-law inflation are given by 
\m
n_s&=&1-{2\over p}, \\
r&=&{16\over p}, 
\n
and then $r=8(1-n_s)$. See the green dashed line in Fig.~\ref{fig:cssi}. We find that the power-law inflation model can fit the data quite well.

Finally we switch to the inflation model with inverse power-law potential \cite{Barrow:1993zq} in which the potential of inflaton field is given by 
\m
V(\phi)=\mu^4 \({M_p\over \phi}\)^n, 
\n
where $\mu$ is an energy scale and $n>0$. The inflation with inverse power-law potential can be ended by some mechanism like that in the hybrid inflation when $\phi_{\rm end}^2=2nN_* M_p^2$, where $N_*$ is roughly the total number of e-folds. The tensor-to-scalar ratio and the spectral index become 
\m
r&=& {4n\over N_*-N}, \\
n_s&=&1-{n-2\over 2(N_*-N)}. 
\n
For $n=2$, $r=8/(N_*-N)$ and $n_s=1$ which implies an HZ spectrum. In this case, $r=0.2$ if $N_*-N=40$. From the above two equations, we obtain $r={8n\over n-2}(1-n_s)$. In the limit of $n\rightarrow \infty$, $r=8(1-n_s)$ which is the same as that in power-law inflation. Anyway, the inflation model with inverse power-law potential can fit the data as well.

To summarize, since W9 is consistent with almost all of other cosmological observations, we adopt the combinations of B2+W9 and B2+W9+SNLS+BAO+HST to constrain the cosmological parameters in the base $\Lambda$CDM+r model with $n_t=-r/8$. We find that the chaotic inflation model is marginally disfavored at around $2\sigma$ level, but the power-law inflation and the inflation model with inverse power-law potential can fit the data quite well. 
In addition, how to achieve a large amplitude of relic gravitational waves spectrum in string theory is still a big challenge \cite{McAllister:2007bg,Baumann:2009ni}.

After B2 released its data, many inflation models were investigated in the last few weeks. For example, see \cite{Dimopoulos:2014boa,Nakayama:2014koa,Harigaya:2014sua,Harigaya:2014qza,Kehagias:2014wza,Kobayashi:2014jga,Freese:2014nla,Gong:2014cqa,Germani:2014hqa,Kawasaki:2014lqa,Bamba:2014jia,Lyth:2014yya,Ellis:2014rxa,DiBari:2014oja,Feng:2014yja,Hazra:2014jka}. We believe that it is still too early to say which model is correct. It is worthy exploring both the data and theoretical models further in the near future.

\vspace{5mm}
\noindent {\bf Acknowledgments}. 
We acknowledge the use of Planck Legacy Archive, ITP and Lenovo
Shenteng 7000 supercomputer in the Supercomputing Center of CAS
for providing computing resources. This work is supported by the project of Knowledge Innovation Program of Chinese Academy of Science and grants from NSFC (grant NO. 10821504, 11322545 and 11335012).



\end{document}